\documentclass[aps,amssymb,amsmath,amsfonts,twocolumn]{revtex4}
\usepackage{graphicx}
\usepackage{dsfont}
\usepackage{verbatim} 
\usepackage{mathbbol,bm,bbm}

\begin{document}
\title{Product of Ginibre matrices: Fuss-Catalan and Raney distributions}

\medskip

\author{Karol A. Penson$^1$ 
 and Karol {\.Z}yczkowski$^{2,{3}}$ 
\\ 
\medskip
 {\normalsize\itshape $^{{1}}$ Universit{\'e} Paris VI,
  Laboratoire de Physique de la Mati{\`e}re Condens{\'e}e (LPTMC), CNRS
  UMR 7600,  t.13, 5{\`e}me {\'e}t. BC.121, 4, pl. Jussieu, F 75252 Paris Cedex 05, France }\\
{\normalsize\itshape {$^2$
Institute of Physics,
   Jagiellonian University, ul. Reymonta 4, 30-059 Krak{\'o}w, Poland}}\\
 {\normalsize\itshape $^{{3}}$Centrum Fizyki Teoretycznej, Polska Akademia Nauk,
   Al. Lotnik{\'o}w 32/44, 02-668 Warszawa, Poland}\\[2ex]}

\medskip 

\email{penson@lptl.jussieu.fr; karol@tatry.if.uj.edu.pl}

\medskip

 \date{{May 5, 2011}}

\begin{abstract}
Squared singular values of a product of $s$ square random Ginibre
matrices are asymptotically characterized by probability distributions
$P_s(x)$, such that their moments are equal to the Fuss--Catalan numbers
or order $s$.
We find  a representation of the Fuss--Catalan distributions $P_s(x)$ 
in terms of a combination of $s$ hypergeometric functions of the type
${}_sF_{s-1}$. The explicit formula derived here is exact for an arbitrary positive integer $s$ 
and for $s=1$ it reduces to the Marchenko--Pastur distribution.
Using similar techniques, involving Mellin transform and the Meijer
$G$--function, we find exact expressions for the Raney probability distributions,
the moments of which are given by a two parameter
generalization of the Fuss-Catalan numbers. These distributions
 can also be considered as a two parameter generalization of the Wigner semicircle law.
\end{abstract}

\maketitle

\section{Introduction}

\medskip

Random matrices of various ensembles find numerous applications in 
several fields of statistical physics.
In the general class of non-hermitian random matrices an
important role is played by the {\sl Ginibre ensemble} \cite{Gi65}.
A matrix $G$ of size $N$ of such an ensemble consists of $N^2$
independent random complex numbers, drawn according to the
Gaussian distribution with zero mean and a fixed variance  \cite{Me04,Fo10}.
Such matrices are used to describe non-unitary dynamics of
chaotic systems  and open quantum systems \cite{Ha10}.
This ensemble of random matrices can also be used
to analyze the human EEG data \cite{Se03},
for telecommunication applications based on the scattering 
of electromagnetic waves on random obstacles \cite{TV04},
 or in  mathematical finances 
to describe correlation matrices of various stocks \cite{BP,BJN03}.

The spectrum of a non-Hermitian matrix $G$ 
belongs to the complex plane.
Spectral density of random matrices of the suitably normalized
Ginibre ensemble is described by the Girko circular law \cite{Gi84}, 
as in the limit $N \to \infty$  it covers uniformly the unit disk.
The random matrix $W=GG^{\dagger}$,  called a {\sl Wishart matrix},
is positive. Hence its eigenvalues $\lambda_i$, $i=1,\dots, N$,
are real and non-negative. Introducing a rescaled eigenvalue, $x=N\lambda$
one can show 
that in the limit of the large matrix size the spectral density $P(x)$
converges to the {\sl Marchenko--Pastur} (MP) distribution \cite{MP67}.

In general, products of random matrices are a subject of an intensive research
for many years  \cite{CPV93}.
Recent studies on products of Ginibre matrices
concern multiplicative diffusion processes \cite{GJJN03},
correlation matrices used in macroeconomic time series  \cite{BLMM07},
a random matrix approach to quantum chromodynamics \cite{Os04} 
and lattice gauge field theories \cite{LNW08}.
Properties of the complex spectra of products of random Ginibre matrices
were recently analyzed in  \cite{BJW10}.

It is also interesting to study singular values of a product
of $s$ independent Ginibre matrices, $X=G_1\cdots G_s$.
Note that a squared singular value of the product $X$ equals 
the corresponding eigenvalue of the Wishart like matrix $W=XX^{\dagger}$.
For $s=2$, positive random matrices of the form
$W_2=G_1G_2 (G_1 G_2)^{\dagger}$ found their 
applications in finances \cite{BLMM07}. 
Matrices of the form $W_s=G_1 \cdots G_s (G_1 \cdots G_s)^{\dagger}$
for an arbitrary $s$ were used to describe random quantum states
associated with certain graphs \cite{CNZ10} 
and quantum states obtained by orthogonal
measurements in a product of
maximally entangled bases \cite{ZPNC10}.

The corresponding asymptotic level density distribution $P_s(x)$ is
called {\sl Fuss-Catalan distribution} of order $s$,
since its moments are given \cite{armstrong,Ml10,banica-etal}
by the Fuss-Catalan numbers \cite{GKP,Ko08},
 (also called Fuss \footnote{When Leonard Euler, after an eye operation in 1772, 
became almost completely blind, he asked Daniel
Bernoulli in Basel to send a young assistant, well trained in mathematics, 
to him in St.~Petersburg.
It was Nikolaus Fu{\ss} who arrived in St.~Petersburg in May 1773 \cite{Oz}.}
numbers \cite{PS00}). Strangely enough, the Fuss-Catalan numbers
generalize the Catalan numbers, although the work of Fu{\ss} \cite{Fu791}
was done much earlier then the contribution of Catalan \cite{Ca838}.
The Catalan number can be defined
as a number of different bracketing of a product of $n+1$ numbers,
or the number of possible $n$ folding of a 
map which contains $n+1$ pages in a row \cite{Ko68}.

The Fuss--Catalan distribution describes asymptotically statistics
of singular values of the $s$--th power of  
random Ginibre matrices.
This result obtained recently by Alexeev et al. \cite{AGT10}
was derived by estimating the moments of the
distribution of squared singular values of a power $G^s$
of a random matrix and showing that these moments 
converge asymptotically to the Fuss--Catalan numbers.
This is true under rather weak assumptions: 
all entries of the matrix $G$ are independent random variables 
characterized by the zero mean, variance set to unity
and finite fourth moment.

The Fuss--Catalan distribution can be considered as a generalization 
of the MP distribution, which is obtained for $s=1$.
Moreover,  the distribution $P_s(x)$ 
belongs to the class of free Meixner measures \cite{bozejko-bryc},
and in terms of free probability theory
it appears as the {\sl free}
 multiplicative convolution product of $s$ copies of the 
MP distribution \cite{banica-etal,BG09}, 
which is written as $P_s(x)=[P_1(x)]^{\boxtimes s}$.

An explicit form in the case $s=2$  was derived in \cite{PS01} in context of 
construction of generalized coherent states from combinatorial sequences.
The spectral distribution of $P_s(x)$ for a product of an arbitrary number 
of $s$ random Ginibre matrices was recently analyzed by Burda et al. \cite{BJLNS10} 
also in the general case of rectangular matrices. The distribution was expressed
as a solution of a polynomial equation and it was conjectured that 
the finite size effects can be described by a simple multiplicative correction.
Another recent work of Liu at al. \cite{LSW10}
provides an integral representation of the distribution  $P_s(x)$
derived in the case of $s$ square matrices of size $N$, which is assumed to be large.
However, these recent contributions do not provide an explicit 
form of the distribution $P_s(x)$.

The aim of this note is to derive exact and explicit
formulae for the Fuss-Catalan distribution $P_s(x)$,
which can be represented as a combination of $s$
hypergeometric functions.
The derivation is presented in Sec. II, while
some auxiliary information on special functions and the
proof of positivity of  $P_s(x)$ 
are provided in Appendices A and B, respectively.
In section III we discuss a certain two-parameters 
generalization of Fuss-Catalan numbers. As these numbers
quantify generalized Raney sequences \cite{GKP}, 
the corresponding probability measures $W_{p,r}(x)$ will be called
{\sl Raney distributions}. As special cases they include
Marchenko--Pastur distribution, Fuss--Catalan distributions
and the Wigner semicircle law. We 
find exact expressions for Raney distributions
corresponding to integer parameter values in the general case
and provide explicit formulae in the simplest cases
of small values of the integer parameters $1\le r\le p$.

\section{Fuss--Catalan distributions}

For any integer number $s$ one can use the binomial symbol 
to define a sequence of integers
denoted by $FC_s(n)$, 
\begin{equation}
FC_s(n) := \frac{1}{sn+1}\binom{sn+n}{n}  \ .
\label{FCmom}
\end{equation}
Here $n=0,1,\dots$, while $s=1,2,\dots$, and
these numbers are called the {\sl Fuss--Catalan} numbers of order $s$ \cite{GKP}.

 We enumerate some of these sequences below for $n=0,\dots,7$:
\begin{eqnarray} 
FC_1(n) & = & 1,1,2,5,14,42,132,427,\dots \nonumber \\
FC_2(n) & = & 1,1,3,12,55,273,1428,7752,\dots \nonumber \\
FC_3(n) & = & 1,1,4,22,140,969,7084,53820,\dots \nonumber \\
FC_4(n) & = & 1,1,5,35,285,2530,23751,231880,\dots \; . \nonumber 
\label{FCseqen}
\end{eqnarray}
The above sequences are contained in the 
Online Encyclopedia of Integer Sequences (OEIS) \cite{Sloane}
under the labels (A000108), (A001724), (A002293)
and (A002294), respectively.
These  sequences can be considered as a generalization
of the sequence $FC_1(n)$,
 which consists of {\sl Catalan numbers},
$C(n)=\frac{1}{n+1} \binom{2n}{n}$.

We are going to show that for any given $s$ 
there exists a density distribution $P_s(x)$, which satisfies 
\begin{equation}
\int_0^{K_s} x^n P_s(x) dx = FC_s(n), \ \ n=0,1,\dots
\label{densmom}
\end{equation}
where 
\begin{equation}
K_s:=(s+1)^{s+1}/s^s \; .
\label{Ks}
\end{equation}
In other words we are looking for a positive density $P_s(x)$
which satisfies the above infinite system of equations.
As the density turns out to be defined in a finite segment
$[0,K_s]$, the solution of this  
{\sl Hausdorff moment problem} \cite{Ak65}
 associated with the Fuss-Catalan numbers  is unique.
An explicit proof of positivity of $P_s(x)$ 
is provided in Appendix B.

We employ the method of the inverse Mellin transform,
which was previously used to construct explicit solutions
of the Hausdorff moment problem \cite{KPS01,GPHDS}
and to derive explicit form of the L{\'e}vy--stable distributions \cite{PG10}.
The {\sl Mellin transform} ${\cal M}$ of a function $f(x)$ and its inverse
${\cal M}^{-1}$ are defined by a pair of equations
\begin{equation}
f^*(\sigma):= \; {\cal M}[f(x);\sigma] = \int_0^{\infty}
  x^{\sigma-1} f(x) dx
\label{mellin1}
\end{equation}
and
\begin{equation}
f(x):= \; {\cal M}^{-1}[f^*(\sigma);x] 
 = \frac{1}{2\pi i}  \int_{c-i\infty}^{c+i\infty}  x^{-\sigma} f^* (\sigma) d\sigma, 
\label{mellin2}
\end{equation}
with complex $\sigma$.
In (\ref{mellin2}) this variable is integrated over a vertical line in 
the complex plane \cite{FGD95}.
For discussion concerning the role of $c$, see (\cite{FGD95,PBM98,Lu69}). 
Therefore the solution of Eq. (\ref{densmom}) can be obtained by extending
integer variable $n$ to complex $\sigma$  by a substitution $n\to \sigma -1$.
The desired form of the distribution $P_s(x)$ can be formally written as an inverse
Mellin transform, 
\begin{equation}
P_s(x) \ = \ {\cal M}^{-1}[FC_s(\sigma);x] \ .
\label{mellin3}
\end{equation}
To find such a transform we will bring 
the Fuss-Catalan numbers  into a more suitable form.
Representing the binomial symbol in (\ref{FCmom})
by the ratio of Euler's gamma function one obtains 
\begin{equation}
FC_s(\sigma) = 
\frac{\Gamma\bigl[ (s+1)\bigl(\sigma -\frac{s}{s+1}\bigr) \bigr] }
{\Gamma \bigl[ s \bigl( \sigma - \frac{s-2}{s} \bigr) \bigr] 
\;  \Gamma (\sigma) }  \ .
\label{FSbis}
\end{equation}

After applying twice the Gauss--Legendre formula (\ref{multeuler}) 
for multiplication of the argument of 
the gamma function one arrives at 
\begin{eqnarray}
FC_s(\sigma) &= &  \frac{1}{\sqrt{2\pi}}  
 \Bigl[\frac{(s+1)^{s+1}}{s^s}\Bigr]^{\sigma} \times 
 \nonumber \\ 
 & \times & \; 
\frac{s^{s-3/2}}{(s+1)^{s+1/2}} \; 
\Bigl[ \prod_{j=0}^{s-1} 
\frac{\Gamma\bigl(\sigma+ \frac{j-s}{s+1}\bigr)}
     {\Gamma\bigl(\sigma+ \frac{2+j-s}{s}\bigr)}
 \Bigr] .
\label{FStri}
\end{eqnarray}
Obtaining the above form of the FC numbers,
in which a ratio of products of the gamma functions of a shifted argument
 appears, is a key step of our reasoning.
It allows us to represent the inverse Mellin transform of Eq. (\ref{FStri})
as a certain special function.
To see this recall that the {\sl Meijer $G$--function}
of the argument $z$ 
can be defined by the inverse Mellin transform \cite{PBM98}:
\begin{eqnarray}
\label{Meijer}
&& G_{p,q}^{m,n} \Bigl( z \; \bigl|\; {\stackrel{\scriptstyle
                       \alpha_1 \cdots \alpha_p}
                {\scriptstyle \beta_1 \cdots  \beta_q}} \Bigr) = 
 \\
 &&    =   {\cal M}^{-1}
\Bigl[ 
\frac{\prod_{j=1}^{m}\Gamma(\beta_j+\sigma) \prod_{j=1}^{n}\Gamma(1-\alpha_j-\sigma)}
     {\prod_{j=m+1}^{q}\Gamma(1-\beta_j-\sigma)
   \prod_{j=n+1}^{p}\Gamma(\alpha_j+\sigma)}; \; z
 \Bigr] . \nonumber
\end{eqnarray}

This definition involves four lists of parameters,
which can be represented by $p$ complex numbers $\alpha_j$ and 
other $q$ complex numbers $\beta_j$.
Integers  numbers $p$ and $q$ can be equal zero and it s is assumed
that $0\le m\le q$ and $0 \le n \le p$, so that possibly empty products 
in this form are taken to be equal to unity.
A detailed description of the integration contours of the Mellin transform 
(\ref{Meijer}), general properties of the Meijer functions  and its special 
cases can be found in \cite{PBM98}.

\begin{figure}[hbp]
\centering
\includegraphics[width=0.32\textwidth]{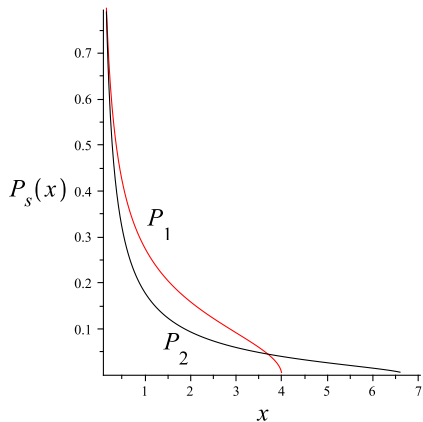}
\caption{ (Color online) Marchenko--Pastur distribution $P_1(x)$
 compared with the Fuss--Catalan distribution $P_2(x)$.
The singularity at $x\to 0$
is of the type  $P_s(x)\sim x^{-s/(s+1)}$.}
\label{fig1}
\end{figure}

\begin{figure}[htbp]
\centering
\includegraphics[width=0.36\textwidth]{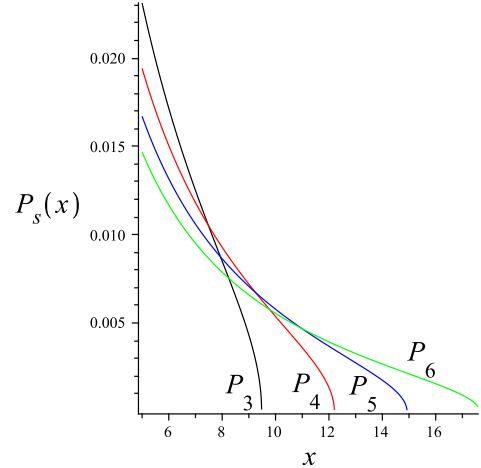}
\caption{ (Color online) Fuss--Catalan distributions $P_s(x)$
plotted for parameter $s$ equal to $3,4,5$ and $6$ are supported on the interval 
$[0,(s+1)^{s+1}/s^s]$. To show behaviour at the right 
side of the support 
the figure is plotted for $x \ge 5$.
}
\label{fig2}
\end{figure}

Direct comparison of expression (\ref{FStri}) for the Fuss Catalan numbers
 and the Mellin transform as the  Meijer $G$--function (\ref{Meijer}) 
allows us to represent the Fuss-Catalan distribution $P_s(x)$ 
by a Meijer $G$--function,
\begin{equation}
P_s(x) =  \frac{1}{\sqrt{2\pi}} \frac{s^{s-3/2}}{(s+1)^{s+1/2}} \;
G_{s,s}^{s,0} \Bigl( z \; \bigl|\; {\stackrel{\scriptstyle
                       \alpha_1 \cdots \alpha_p}
                {\scriptstyle \beta_1 \cdots  \beta_q}} \Bigr)
\label{FCdist1}
\end{equation}
of the argument $z=x s^s (s+1)^{-(s+1)}$.
Looking at the range of the parameter  $j$ in (\ref{FStri})
we see that the numbers of parameters of the Meijer $G$--function 
have to be set to $n=0$,  $p=s$ and $m=q=s$.
Hence this function involves $2s$ parameters, which  read  
$\alpha_j=(1+j-s)/s$ and $\beta_j=(j-1-s)/(s+1)$ for $j=1,\dots, s$.

As there are only two products of gamma functions
in (\ref{FStri}), in contrast to four in 
(\ref{Meijer}), which is equivalent to setting $n=0$ and $m=q$,
a further simplification of the above formula is possible.
In this very case the Meijer $G$--function
can be written as a combination of hypergeometric functions
of the same argument $z$ -- see Eq. (5.2.11) p. 146 of \cite{Lu69}.

Let  ${}_pF_q\Bigl( \bigl[ \{a_j\}_{j=1}^p \bigr] , \; 
                    \bigl[ \{b_j\}_{j=1}^q \bigr] ; \; x\Bigr)$
denote the hypergeometric function \cite{MOT} of the type ${}_pF_q$
with $p$ 'upper' parameters $a_j$ and $q$ 'lower' parameters $b_j$
of the argument $x$. The symbol 
$\{a_i\}_{i=1}^r $ represents the list of $r$ elements, $a_1,\dots a_r$.
Then formula (\ref{FCdist1}) for the Fuss--Catalan 
distribution can be rewritten as

\begin{widetext}

\begin{equation}
\label{eq:FCs}
P_s(x) =
 \sum_{k=1}^s  \Lambda_{k,s}\; x^{\frac{k}{s+1}-1}\; 
{}_sF_{s-1}\Bigl( \Bigl[ \Bigl\{1-\frac{1+j}{s} +\frac{k}{s+1} \Bigr\}_{j=1}^s \Bigr] , \;
 \Bigl[ \Bigl\{1+\frac{k-j}{s+1} \Bigr\}_{j=1}^{k-1} ,
      \Bigl\{1+\frac{k-j}{s+1} \Bigr\}_{j=k+1}^{s} \Bigr] ; \;
         \frac{s^s}{(s+1)^{s+1}} x 
                   \Bigr) \ .
\end{equation}
where the coefficients $\Lambda_{k,s}$ read for $k=1,2, \dots, s$
\begin{equation}
\label{eq:FCs2}
\Lambda_{k,s} := s^{-3/2} \sqrt{\frac{s+1}{2\pi}} 
 \Bigl(\frac{s^{s/(s+1)}}{s+1}\Bigr)^{k}  \
 \frac{\Bigl[ \prod_{j=1}^{k-1} \Gamma\bigl(\frac{j-k}{s+1}\bigr) \Bigr]
       \Bigl[ \prod_{j=k+1}^{s} \Gamma\bigl(\frac{j-k}{s+1}\bigr) \Bigr] }
  { \prod_{j=1}^{s} \Gamma\bigl( \frac{j+1}{s} - \frac{k}{s+1}\bigr) } \ .
\end{equation}

\end{widetext}

This formula, along with Eq.(\ref{raney2}) below, constitutes 
the key result of the present note.
It gives an exact result for the Fuss--Catalan distribution $P_s(x)$
for an arbitrary natural $s$.
The FC distribution describes the density of squared singular 
values of a product of $s$ independent square Ginibre matrices
in the limit of large matrix size.

The convergence conditions of the hypergeometric series 
${}_sF_{s-1}$ immediately yield the support of $P_s(x)$
which is equal to $[0,(s+1)^{s+1}/s^{s}]$.
For small values of $x$ the distribution behaves as $x^{-s/(s+1)}$.
It is conforting to see that in the simplest case $s=1$ the above 
complicated form reduces indeed  to
the Marchenko--Pastur distribution, 
\begin{equation}
P_1(x) = \frac{1}{\pi \sqrt{x}} 
\; {}_1F_0 \Bigl( [-\frac{1}{2}], [ \ ] ; \; \frac{1}{4}x \Bigr) 
=\frac{\sqrt{1-x/4}}{\pi \sqrt{x}}\; .
\label{pi1bis}
\end{equation}
Furthermore, the distribution $P_2(x)$, shown in Fig. \ref{fig1},
\begin{eqnarray}
P_2(x) =&& \frac{\sqrt{3}}{2\pi x^{2/3}} 
\;{}_2F_1 \Bigl( \bigl[-\frac{1}{6},\frac{1}{3}\bigr], \;
                 \bigl[\frac{2}{3} \bigr] ; \; \frac{4}{27}x \Bigr)
     \ \ \ \  \ \ \ \ \nonumber  \\ 
 &&  
 - \frac{\sqrt{3}}{6\pi x^{1/3}} 
\;{}_2F_1 \Bigl( \bigl[\frac{1}{6},\frac{2}{3}\bigr] , \; 
   \bigl[\frac{4}{3}\bigr]; \; \frac{4}{27}x \Bigr)
\label{p21bis}
\end{eqnarray}
is equivalent to the form,
\begin{equation}
\!\!
P_2(x) =  \frac{\sqrt[3]{2} \sqrt{3}}{12 \pi} \;
 \frac{\bigl[\sqrt[3]{2} \left(27 + 3\sqrt{81-12x} \right)^{\frac{2}{3}} -
   6\sqrt[3]{x}\bigr] } {x^{\frac{2}{3}}
     \left(27 + 3\sqrt{81-12x} \right)^{\frac{1}{3}}},  
\label{pi2}
\end{equation}
valid for $ x \in [0,27/4]$ and
 obtained first in \cite{PS01} in context of 
construction of generalized coherent states from combinatorial sequences.
The distribution $P_3(x)$, plotted in Fig. \ref{fig2},
 is given by a sum of three terms,
\begin{eqnarray}
P_3(x) = \!\!
&& \frac{1}{\sqrt{2} \pi x^{3/4}}
\;{}_3F_2 \Bigl( \bigl[-\frac{1}{12},\frac{1}{4}, \frac{7}{12}\bigr], \;
 \bigl[\frac{1}{2},\frac{3}{4}\bigr]; \;  \frac{27}{256}x \Bigr)
     \ \ \ \  \ \ \ \ \nonumber  \\ 
 && 
\!\!
 - \frac{1}{4 \pi x^{1/2}}
\;{}_3F_2 \Bigl( \bigl[ \frac{1}{6},\frac{1}{2}, \frac{5}{6}\bigr], \;
\bigl[ \frac{3}{4},\frac{5}{4}\bigr]; \; \frac{27}{256}x \Bigr)
     \ \ \ \  \ \ \ \ \nonumber  \\ 
&& 
\!\!
 - \frac{\sqrt{2}}{64 \pi x^{1/4}}
\;{}_3F_2 \Bigl( \bigl[\frac{5}{12},\frac{3}{4}, \frac{13}{12}\bigl], \;
\bigl[\frac{5}{4},\frac{3}{2}\bigr]; \; \frac{27}{256}x \Bigr) .
\label{pi3}
\end{eqnarray}
In Fig. 2 we present the  Fuss--Catalan distributions  
$P_s(x)$ for $s=3,4,5$ and $6$.

\section{Raney distributions}
\label{sec_Raney}

The Fuss--Catalan numbers $FC_s(n)$ defined in (\ref{FCmom}) can be 
considered as a special cases of a larger family of sequences,
\begin{equation}
R_{p,r}(n) := \frac{r}{pn+r}\binom{pn+r}{n}  
\label{Ran1}
\end{equation}
defined for $n=0,1,\dots$. 
Here $p$ and $r$ are treated as integer parameters, 
$p\ge 2$, $r=1,2,\dots$ .
Setting $r=1$ and $p=s+1$ we have
$\frac{1}{np+1}\binom{np+1}{n}=\frac{1}{n(p-1)+1}\binom{np}{n}$,
so the numbers $R_{s+1,1}(n)$ are equal to $FC_s(n)$.
Further relations involving the sequences $R_{p,r}(n)$ are
\begin{equation} 
R_{p+1,p+1}(n) = FC_p(n+1),
\label{Ranrel1}
\end{equation}
and
\begin{equation} 
R_{p,p}(n) = R_{p,1}(n+1) ,
\label{Ranrel2}
\end{equation}
which can be verified directly from their definitions (\ref{FCmom}) and (\ref{Ran1}).

The Raney lemma \cite{Ra60} implies
that the number of the Raney sequences of order $p$ and length $pn+1$,
for which all partial sums are positive,
is given by the Fuss--Catalan numbers $FC_{p-1}(n)=R_{p,1}(n)$.
Furthermore, as  the number of positive generalized Raney sequences 
is equal to $R_{p,r}(n)$ \cite{GKP}
we will refer to $R_{p,r}(n)$ defined in (\ref{Ran1}) 
as {\sl Raney numbers}. These numbers 
appear as coefficients in a generalized binomial series \cite{GKP}.
Some representative examples of sequences $R_{p,r}(n)$,
for $n=0,1,\dots,7$ are quoted below
together with their OEIS labels \cite{Sloane},
\begin{eqnarray} 
R_{4,2}(n) & = & 1,2,9,52,340,2394,17710,135720, ... (A069271) \nonumber \\
R_{5,2}(n) & = & 1,2,11,80,665,5980,56637,556512, ... (A118969), \nonumber 
\label{ranseq1}
\end{eqnarray}
whereas two following sequences are not represented in OEIS:
\begin{eqnarray} 
R_{4,5}(n) & = & 1,5,30,200,1425,10626,81900,647280, ... \; ,\nonumber \\
R_{6,3}(n) & = & 1,3,21,190,1950,21576,250971,3025308,... \; . \nonumber 
\label{ranseq2}
\end{eqnarray}

\medskip 
In a recent work M{\l}otkowski \cite{Ml10}
has shown that the sequence (\ref{Ran1})
describes moments of a probability measure $\mu_{p,r}$ with a compact support
contained in $[0,\infty)$,
if point $(r,p)$ which determines parameters of the Raney numbers
belongs to the set $\Sigma$ defined by inequalities
$p\ge 0$ and $0 < r \le p$.
Note that the point $(1,1)$ implies a constant sequence of moments,
$R_{1,1}(n)=1$,  which represents a singular, Dirac delta measure, 
$\mu_{1,1}=\delta(x-1)$.

In the case the measure $\mu_{p,r}$
is represented by a density we will denote it by $W_{p,r}(x)$. 
Setting $r$ to one one gets the Fuss--Catalan numbers,
which implies that $W_{2,1}(x)$ represents the Marchenko--Pastur distribution,
while $W_{s+1,1}(x)$ reduces to the
Fuss--Catalan probability density $P_s(x)$.

In general parameters $p$ and $r$ can be taken to be real 
and then the moments of the measure are expressed by the gamma functions,
\begin{equation}
\int x^n \mu_{p,r}(x) dx = 
   \frac{r}{np+r} 
\frac{\Gamma(np+r+1)} {\Gamma(n+1) \Gamma(np+r-n+1)} , 
\label{Ran2}
\end{equation}
where the integration covers entire support of the measure $\mu_{p,r}$.
For $1 \le r \le p$ the distribution $W_{p,r}(x)$ is a positive function, 
see \cite{Ml10} and Appendix B.

The corresponding distribution $W_{p,r}(x)$ can be written implicitly
by its $S$-transform, which allowed M{\l}otkowski 
to establish relations between various Raney distributions \cite{Ml10},
listed in Appendix C.
For a precise definition of the $S$--transform
(or the free multiplicative transform) see Eq. 4.9 in \cite{Ml10}.
In spite of these concrete results 
an explicit form of the Raney distribution $W_{p,r}(x)$ 
has not appeared in the literature so far.

Making use of the inverse Mellin transform and the Meijer function
we can generalize results of the previous
section and obtain explicit expressions
for the Raney distributions $W_{p,r}(x)$,
which correspond to integer values of the parameters $p$ and $r$.

Repeating steps analogous to Eqs.(\ref{FSbis}) -- (\ref{FCdist1})
we can represent distribution 
$W_{p,r}(x)$ in terms of the Meijer $G$--function.
The explicit expression generalizing Eq.(\ref{FCdist1})
reads 
\begin{equation}
W_{p,r}(x) = \frac{r}{\sqrt{2 \pi}} \frac{p^{r-p-1/2} } {(p-1)^{r-p+3/2}}
\; G_{p,p}^{p,0} \Bigl( z \; \bigl|\; {\stackrel{\scriptstyle
                       \alpha_1 \cdots \alpha_p}
                {\scriptstyle \beta_1 \cdots  \beta_q}} \Bigr) ,
\label{raney}
\end{equation}
where the argument of the function is $z=x(p-1)^{p-1}/p^p$,
while its parameters are $\alpha_1=0$, $\alpha_j=(r-p+j)/(p-1)$ for
$j=2,\dots, p$
and  $\beta_j=(r-p-1+j)/p$ for $j=1,\dots, p$.
In analogy to (\ref{eq:FCs}) this relation can be 
represented by the following sum consisting of of $p$ terms:
\begin{widetext}

\begin{equation}
\!\!
W_{p,r}(x) = 
\sum_{j=1}^{p} \Omega (p,r;j) x^{\frac{r-1+j}{p}-1}\; 
{}_pF_{p-1}\Bigl( \Bigl[
1+\beta_j,
 \Bigl\{1+\beta_j-\alpha_{i}\Bigr\}_{i=2}^p \Bigr] , \;
\Bigl[\Bigl\{1+\frac{j-i_2}{p} \Bigr\}_{i_2=1}^{j-1}, 
      \Bigl\{1+\frac{j-i_3}{p} \Bigr\}_{i_3=j+1}^{p} \Bigr] ; \;
         \frac{(p-1)^{p-1}}{p^p} x   \Bigr) \; ,
\label{raney2}
\end{equation}
where ${}_pF_{p-1}$ is the hypergeometric function and the 
numerical coefficients  $\Omega(p,r;j)$, for $j=1,2, \dots, p$
read
\begin{equation}
\label{eq:raney3}
\Omega(p,r;j) := \frac{r}{\sqrt{2\pi}}
\frac{p^{r-p-1/2}}{(p-1)^{r-p+3/2}}
  \Bigl(\frac{(p-1)^{p-1}}{p^p}\Bigr)^{\frac{r-p-1+j}{p}}  \;
\frac{1}{ \Gamma \bigl( \frac{p-r+1-j}{p} \bigr)}\;
 \frac{\Bigl[ \prod_{i_1=1}^{j-1} \Gamma\bigl(\frac{i_i-j}{p}\bigr) \Bigr]
       \Bigl[ \prod_{i_2=1}^{p-j} \Gamma\bigl(\frac{i_2}{p}\bigr) \Bigr] }
  { \prod_{i_3=2}^{p} \Gamma\bigl(\frac{r-p+i_3}{p-1} - \frac{r-p-1+j}{p}\bigr) } \ .
\end{equation}
\end{widetext}

Convergence properties of the hypergeometric function
imply that the Raney distribution $W_{p,r}(x)$ for integer values of
its parameters is supported in the interval $[0,K_{p-1}]$,
where $K_s$ is given in Eq. (\ref{Ks}).
Formula (\ref{raney2}) implies that for $p>r$ the distribution $W_{p,r}(x)$
diplays a singularity for small $x$ of the type $x^{-(p-r)/p}$.
For $p=r$ the 'diagonal' Raney distributions behave for small arguments
 as $W_{p,p}(x) \sim x^{1/p}$.

It is helpful to add some clarifying remarks concerning the
key formula (\ref{raney2}).
We draw attention to the fact that some simplifications will
always appear for two reasons:

a) firstly, one parameter form the 'upper' list of the parameters
  will be always equal to a parameter from the 'lower' list.
  Consider, for instance, the case $p=4$ and $r=2$. Then the value
   of the first parameter in the 'upper' list becomes $1+(j-3)/4$
and it cancels with the value of $i_3=3$ of the first sequence 
of the 'lower' parameters. One can demonstrate that
a similar cancellation effect takes place for any pair $(p,r)$.
Therefore, the hypergeometric function ${}_pF_{p-1}$ in (\ref{raney2})
effectively reduces to  ${}_{p-1}F_{p-2}$.

b) secondly, we see from Eq.(\ref{eq:raney3})
that the coefficient $\Omega(p,r;j)$ vanishes for $j=p+1-r$, due to
the presence of the first gamma function in denominator. Therefore the
sum in Eq. (\ref{raney2}) involves $(p-1)$ terms with different hypergeometric functions
${}_{p-1}F_{p-2}$. Then it can be explicitly verified that from the equality
between the numbers $R_{s+1,1}(n)=FC_s(n)$, the corresponding equality
between the probability distributions
\begin{equation}
W_{s+1,1}(x) \ = \ P_s(x)
\label{raney6}
\end{equation}
follows. In particular the following identity between the coefficients holds,
$\Omega(s+1,1;j)=\Lambda_{j,s}$, whose demonstration 
is rather tedious and will not be reproduced here. 

\begin{figure}[htbp]
\centering
\includegraphics[width=0.35\textwidth]{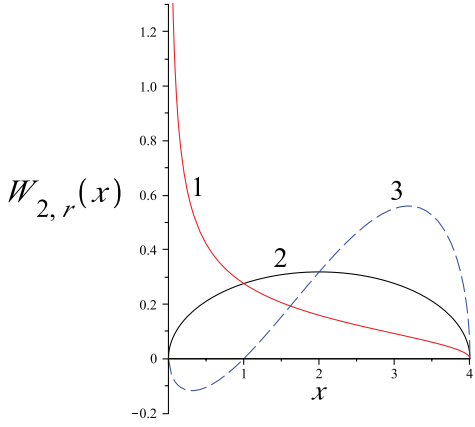}
\caption{ (Color online) 
 Raney distributions $W_{2,r}(x)$ with values of the parameter $r$
labeling each curve. For $r=1$ it reduces to Marchenko-Pastur distribution $P_1(x)$,
while a semicircle law is obtained for $r=2$. For $r=3$ the function
 represented by dashed line is not positive, which is implied by $p<r$.
}
\label{fip2}
\end{figure}

With the two provisos explained under items a) and b) above,
Eq.(\ref{raney2}) will be used to obtain explicit forms
of distributions $W_{p,r}(x)$ for small values of $r$ and $p$.
In particular, the case $W_{2,2}(x)$
reduces to the celebrated semi-circular law \cite{Me04}:
\begin{equation}
\! 
W_{2,2}(x) = \frac{\sqrt{x}}{\pi} 
\; {}_1F_0 \Bigl([-\frac{1}{2}], [ \ ]; \; \frac{x}{4} \Bigr) 
= \frac{1}{2\pi} 
 \sqrt{x(4-x)}  .
\label{R22}
\end{equation}
The above semicircle is centered at $x=2$,
while the Wigner semicircle centered at $x=0$
is used in random matrix theory
to describe the asymptotic level density of
random hermitian matrices form Gaussian ensembles \cite{Me04,Fo10}.
The Raney distributions $W_{2,r}(x)$ are plotted in Fig. \ref{fip2}
for $r=1$ and $r=2$. For comparison we have also plotted the function
$W_{2,3}(x)$ furnished by Eq. (\ref{raney2}). According to the results of M{\l}otkowski
\cite{Ml10} and our Appendix B
 this function is not positive in its domain, 
so it does not represent a probability distribution.
Here we obtain it explicitly.

\begin{figure}[htbp]
\centering
\includegraphics[width=0.35\textwidth]{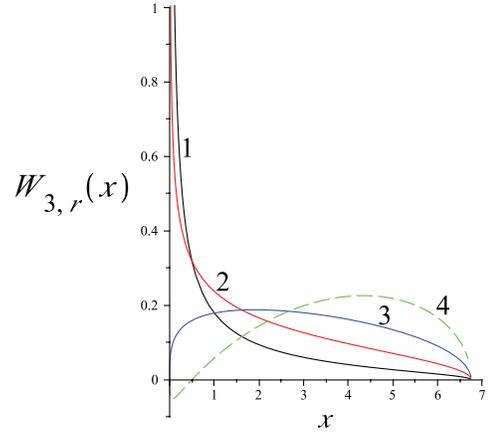}
\caption{ (Color online)
As in Fig.\ref{fip2} for Raney distributions $W_{3,r}(x)$
supported in $[0,6\frac{3}{4}]$.
The case $W_{3,4}(x)$ (dashed line) is not a probability measure.
Each curve is labelled by the value of $r$.
}
\label{fip3}
\end{figure}

It is easy to observe that the above semicircle distribution is related
with the Marchenko-Pastur distributions $P_1(x)$ 
by a relation $W_{2,2}(x)=xP_1(x)$. This is a special case
of a more general relation involving the "diagonal"
Raney distributions with $r=p$ and Fuss-Catalan distribution,
\begin{equation}
W_{p,p}(x) = x\;  W_{p,1}(x) =x P_{p-1}(x).
\label{Rpp}
\end{equation}
This result,  established first in \cite{Ml10},
follows also naturally from Eqs.(\ref{raney}) and (\ref{raney2}).
Thus for $p=3$ one has
\begin{eqnarray}
W_{3,1}(x) = P_2(x),  \ \  {\rm and} \ \
W_{3,3}(x) = x \; P_2(x) ,
\label{W33}
\end{eqnarray}
where the Fuss-Catalan distribution $P_2(x)$ is given in 
(\ref{p21bis}). Due to an equivalent expression (\ref{pi2})
the distribution $W_{3,3}(x)$ can also be expressed in terms of elementary
functions.
The intermediate Raney distribution, corresponding to $r=2$, 
\begin{eqnarray}
W_{3,2}(x) =&& \frac{\sqrt{3}} {2\pi x^{1/3}} 
\;{}_2F_1 \Bigl( \bigl[ -\frac{1}{3},\frac{1}{6}\bigr], \; 
  \bigl[ \frac{1}{3}\bigr] ; \; \frac{4}{27}x \Bigr)
     \ \ \ \  \ \ \ \ \nonumber  \\ 
 &&  
 - \frac{\sqrt{3} x^{1/3}}{18\pi} 
\;{}_2F_1 \Bigl( \bigl[\frac{1}{3},\frac{5}{6}\bigr] , \; 
                 \bigl[\frac{5}{3}\bigr] ; \; \frac{4}{27}x \Bigr)
\label{W32}
\end{eqnarray}
 is shown in Fig. \ref{fip3}.
In a close analogy to Eq. (\ref{pi2})
this distribution enjoys 
a similar representation in terms of elementary functions, 
\begin{equation}
\!\!
W_{3,2}(x) =  \frac{\sqrt{3} \sqrt[3]{2}}{36 \pi} \;
 \frac{\bigl[ \left(27 + 3\sqrt{81-12x} \right)^{\frac{4}{3}} -
   18 \sqrt[3]{2} x^{\frac{2}{3}} \bigr] }
    { x^{\frac{1}{3}}
     \left(27 + 3\sqrt{81-12x} \right)^{\frac{2}{3}}},  
\label{W32b}
\end{equation}
Observe that the distribution $W_{3,1}(x)$,  $W_{3,2}(x)$ and  $W_{3,3}(x)$
behave for small $x$ as $x^{-2/3}$, $x^{-1/3}$ and $x^{1/3}$, respectively.

\begin{figure}[htbp]
\centering
\includegraphics[width=0.35\textwidth]{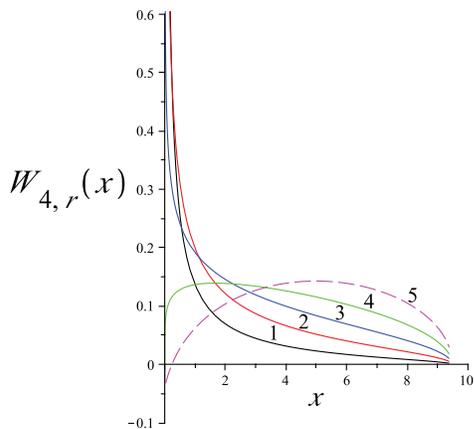}
\caption{ (Color online) As in Fig. \ref{fip2} for Raney distributions $W_{4,r}(x)$.
Dashed line corresponds to a quasi-measure 
$W_{4,5}(x)$, which is not positive.
}
\label{fip4}
\end{figure}

Let us now discuss the case $p=4$ illustrated in Fig. \ref{fip4}.
Due to relations (\ref{Rpp}) one has
\begin{eqnarray}
W_{4,1}(x) = P_3(x),  \ \  {\rm and} \ \
W_{4,4}(x) = x \; P_3(x) ,
\label{W44}
\end{eqnarray}
where the Fuss-Catalan distribution $P_3(x)$ is given in 
(\ref{pi3}).
Two intermediate Raney distributions have a similar form
\begin{eqnarray}
W_{4,2}(x) = \!\!
&& \frac{1}{\pi x^{1/2}}
\;{}_3F_2 \Bigl( \bigl[-\frac{1}{6},\frac{1}{6}, \frac{1}{2}\bigr], \;
 \bigl[\frac{1}{4},\frac{3}{4}\bigr]; \;  \frac{27}{256}x \Bigr)
     \ \ \ \  \ \ \ \ \nonumber  \\ 
 && 
\!\!
 - \frac{\sqrt{2}}{4 \pi x^{1/4}}
\;{}_3F_2 \Bigl( \bigl[ \frac{1}{12},\frac{5}{12}, \frac{3}{4}\bigr], \;
\bigl[ \frac{1}{2},\frac{5}{4}\bigr]; \; \frac{27}{256}x \Bigr)
     \ \ \ \  \ \ \ \ \nonumber  \\ 
&& 
\!\!
 - \frac{\sqrt{2} x^{1/4}} {128 \pi}
\;{}_3F_2 \Bigl( \bigl[\frac{7}{12},\frac{11}{12}, \frac{5}{4}\bigl], \;
\bigl[\frac{3}{2},\frac{7}{4}\bigr]; \; \frac{27}{256}x \Bigr) ,
\label{W42}
\end{eqnarray}
and
\begin{eqnarray}
W_{4,3}(x) = \!\!
&& \frac{1}{\sqrt{2} \pi x^{1/4}}
\;{}_3F_2 \Bigl( \bigl[-\frac{1}{4},\frac{1}{12}, \frac{5}{12}\bigr], \;
 \bigl[\frac{1}{4},\frac{1}{2}\bigr]; \;  \frac{27}{256}x \Bigr)
\nonumber  \\ 
 && 
\!\!
 - \frac{3\sqrt{2}x^{1/4} }{64 \pi}
\;{}_3F_2 \Bigl( \bigl[ \frac{1}{4},\frac{7}{12}, \frac{11}{12}\bigr], \;
\bigl[ \frac{3}{4},\frac{3}{2}\bigr]; \; \frac{27}{256}x \Bigr)
   \nonumber  \\ 
&& 
\!\!
 - \frac{ x^{1/2}} {32 \pi}
\;{}_3F_2 \Bigl( \bigl[\frac{1}{2},\frac{5}{6}, \frac{7}{6}\bigl], \;
\bigl[\frac{5}{4},\frac{7}{12}\bigr]; \; \frac{27}{256}x \Bigr) .
\label{W43}
\end{eqnarray}
The general formula (\ref{raney2}) allows us to obtain an explicit
form of the functions $W_{4,5}(x)$, which is not positive
and it does not represent a probability distribution,
see the dashed curve on Fig.5.

\begin{figure}[htbp]
\centering
\includegraphics[width=0.35\textwidth]{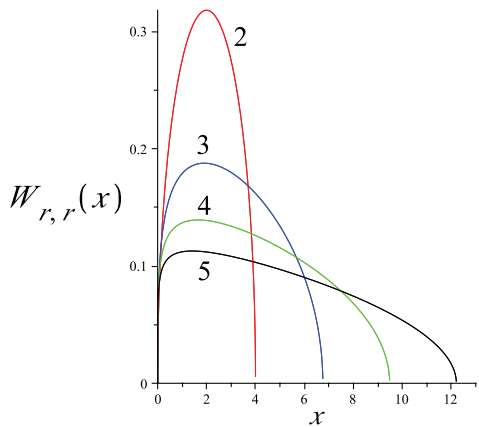}
\caption{ (Color online) 'Diagonal' Raney distributions $W_{r,r}(x)$
form a generalization of the semicircle law $W_{2,2}(x)$.
For small $x$ they behave as $x^{1/r}$.
Value of $r$ labels each curve.
}
\label{fiprr}
\end{figure}

\begin{figure}[htbp]
\centering
\includegraphics[width=0.34\textwidth]{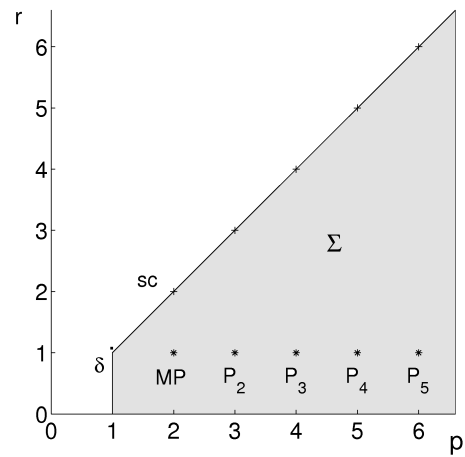}
\caption{Raney distributions in the parameter space $(p,r)$:
Marchenko--Pastur distribution $P_1$
is represented by the point $(2,1)$,
while Fuss-Catalan  distribution $P_s$
is represented by $(*)$ at $(s+1,1)$.
The semicircle  (sc) law corresponds to $(2,2)$
and the 'diagonal' Raney distributions $(+)$ to $(p,p)$. 
As shown in Ref. \cite{Ml10} 
any point in the shaded set $\Sigma$
corresponds to a probability measure with a compact support in $[0,\infty)$.}
\label{figparam}
\end{figure}

Figure \ref{fiprr} presents the Raney distributions $W_{p,r}(x)$
in the "diagonal" case, $r=p$.
As the distribution  $W_{2,2}(x)$ represents Eq.(\ref{R22}), 
the diagonal Raney distributions, $W_{r,r}(x)$,
can  thus by considered as a generalization of the
semicircular law: they are defined for $x\in (0,K_{r-1})$
where $K_s=(s+1)^{s+1}/s^s$,
they are equal to zero at both ends of the domain and
they are characterized by a single maximum.
However, of $r>2$ the functions $W_{r,r}(x)$ are not symmetric
anymore, compare Fig. \ref{fiprr}.
A plot of the parameter space $(p,r)$
in which these distributions are marked 
together with the  Fuss-Catalan distributions
is presented in Fig. \ref{figparam}.

\section{Concluding remarks}
\label{sec_concl}

In this work we obtained an explicit  form of the Fuss-Catalan distribution
$P_s(x)$. Obtained result is exact for an arbitrary $s$
and it allows for a simple use of these probability distributions.
Results derived are relevant from the point of view of statistical physics
as they describe asymptotic level density of  normalized positive random matrix
of the product form $X=(G_1 \cdots G_s)(G_1 \cdots G_s)^{\dagger}$,
where $G_1, \dots ,G_s$ denote $s$ independent random matrices
from the complex Ginibre ensemble. The variable $x=N\lambda$
denotes the rescaled eigenvalue $\lambda$ of random 
matrix $X$ of size $N$.

It should be emphasized here that 
the Marchenko--Pastur and Fuss--Catalan distributions 
describe the level density of Wishart like random matrices 
in the limiting case $N\to \infty$ only.
In practice, for any fixed $N$ the finite size effects 
occur. As discussed by Blaizot and Nowak \cite{BN09,BN10}
the finite $N$ effects are related with the diffraction phenomena, 
while the large $N$ limit of the random matrix theory
may be compared with the geometric limit of the 
wave optics or the semiclassical limit of the quantum theory. 
An explicit description of finite $N$ corrections
to the spectral density of Wishart matrices
obtained from products of Ginibre matrices
is provided by Burda et al. \cite{BJLNS10,BJLNS11}.

A simple argument put forward in \cite{ZPNC10} (see Appendix B therein)
implies that the same FC distributions describe also the spectral density 
of a product of $s$ random matrices taken from the {\sl real Ginibre} ensemble
\cite{LS91}. 
The case of a product of two real matrices, 
recently studied in context of quantum chromodynamics \cite{APS10},
is also important in applications in econophysics,
where one uses a product  of two real correlation matrices \cite{BLMM07}.

The Catalan numbers are a special case of a more general,
one parameter family of Fuss-Catalan numbers, which from a subset
of two-parameter family of Raney numbers. In the same way,
the Marchenko--Pastur distribution $P_1(x)$ is a special case
of the Fuss-Catalan distributions, which in turn belong to the
two parameter family of Raney distributions $W_{p,r}(x)$.
This wide class of probability distribution
includes e.g. the Dirac delta,  $\delta(x-1)$
and the semicircle law, $W_{2,2}(x)$.
Applying the inverse Mellin transform 
for integer parameter values of the parameters
we found an explicit exact representation
of the Raney distributions in terms of
the hypergeometric function.
For any $r=1,2,\dots, p$ the Raney 
distribution $W_{p,r}(x)$ is supported in the interval
$[0,p^p/(p-1)^{p-1}]$. For small $x$ the
distribution behaves as

\begin{equation}
W_{p,r}(x) \ \sim \  
\begin{cases} x^{-\frac{p-r}{p}}   & \textrm{if \ $r < p$},\\
              x^{ \frac{1}{p}}   & \textrm{if \ $r=  p$} .
\end{cases}
\label{singul}
\end{equation}
The Raney numbers  (\ref{Ran1})
imply that the mean value of the Raney distribution
$W_{p,r}(x)$ is equal to $r$,
while the second moment reads $r(2p+r-1)/2$.

Let us conclude the paper with the following remark.
The Fuss--Catalan distribution describe statistical
properties of singular values of
products of random matrices of the Ginibre ensemble.
It is then natural to ask, whether there exist
any ensembles of random matrices, such that
their squared singular values
can be described by the Raney distributions.

\bigskip

Acknowledgments.
It is a pleasure to thank G.~Akemann, Z.~Burda, 
M.~Bo{\.z}ejko, B.~Collins, K.~G{\'o}rska,
A.~Horzela, W.~M{\l}otkowski, I.~Nechita and  A.~M.~Nowak
for fruitful discussions and helpful remarks.
We are thankful to both referees for useful comments.
Financial support by the Transregio-12 project 
of the Deutsche Forschungsgemeinschaft and
the grant number  N N202~090239 of
Polish Ministry of Science and Higher Education
is gratefully acknowledged. The authors acknowledge 
support from Agence Nationale de la Recherche (Paris, France)
under program PHYSCOMB No. ANR-08-BLAN-0243-2.
\appendix

\section{Some information on special functions}
\label{append1}

Special functions of interest for this note are
related to the Euler gamma function, which admits a
following integral representation, 
$\Gamma(z) = \int\limits_0^{+\infty}  t^{z-1}\,e^{-t}\, dt.$
Integrating by parts we see that $ \Gamma(z+1)=z\cdot\Gamma(z)$.
For an integer argument the Euler function is
given by factorial,  $\Gamma(n+1)=n!$\;.
The Gauss--Legendre formula allows one to compute 
the Euler gamma function of a multiple of argument \cite{MOT}, 
\begin{equation}
\Gamma(kz)=(2\pi)^{(1-k)/2} \; k^{kz-1/2} \; 
\prod_{j=0}^{k-1} \Gamma\bigl(z +\frac{j}{k}\bigr)
\label{multeuler} 
\end{equation}
for $k=1,2,3,\dots$ and $z \ne 0, -1,-2,\dots \; .$

The  generalized hypergeometric series 
is a series in which the ratio of successive coefficients indexed 
by $n$ is a rational function of $n$. It can be defined as  
\begin{equation}
\! \! {}_pF_q\bigl(
 [\{a_j\}_{j=1}^p], [\{b_j\}_{j=1}^q] \; ;z) 
:= \sum_{n=0}^\infty \frac{(a_1)_n\dots(a_p)_n}{(b_1)_n\dots(b_q)_n} \, \frac {z^n} {n!} ,
\label{hyperFs}
\end{equation}
where we use the {\sl Pochhammer symbol}, 
defined by 
$(a)_n=a(a+1)(a+2)...(a+n-1)$ and $(a)_0 = 1$.

The series (\ref{hyperFs}), if convergent, defines a 
{\sl generalized hypergeometric function}, which
may then be defined over a wider domain of the argument by analytic continuation. 

Note that for $s=2$ the expression (\ref{eq:FCs}) involves the Gauss 
(ordinary)  hypergeometric function 
\begin{equation}
{}_2F_1( [a,b], [c];\; z) = \sum_{n=0}^\infty \frac{(a)_n(b)_n}{(c)_n} \, 
\frac {z^n} {n!} , 
\label{hyperF2}
\end{equation}
that includes many other special functions as special or limiting cases.

\section{Positivity of the distributions\\
 $P_s(x)$ and $W_{p,r}(x)$}
\label{append2}
For completeness we prove in this appendix that for any integer $s$
the distribution $P_s(x)$ is positive for $x\in (0,K_s)$,
where $K_s=(s+1)^{s+1}/s^s$.
Eq. (\ref{mellin3}) implies  that $P_s(x)$ is given as the inverse
Mellin transform of Eq. (\ref{FStri}).
We are going to use the convolution property for two Mellin transforms,
${\cal M}[f(x);\sigma] =f^*(\sigma)$ and
${\cal M}[g(x);\sigma] =g^*(\sigma)$, which reads \cite{KPS01,GPHDS}
\begin{eqnarray}
 {\cal M}^{-1}[f^*(\sigma)g^*(\sigma);x]  
 =  \int_{0}^{\infty} f\bigl(\frac{x}{t}\bigr) g(t) \frac{dt}{t} \nonumber \\
 =  \int_{0}^{\infty} g\bigl(\frac{x}{t}\bigr) f(t) \frac{dt}{t} .
\label{mellconv}
\end{eqnarray}
If $x>0$ and both functions $f(x)$ and $g(x)$ are positive
then  
their Mellin convolution defined by the integrals  (\ref{mellconv})
conserves positivity.

Consider now, for a given $j=0,1,\dots, (s-1)$, the individual term
in the product in Eq.(\ref{FStri}).
 Its Mellin transform will satisfy, 
due to formula 8.4.2.3, p.631 in \cite{PBM98},
the following equality 
\begin{equation}
 {\cal M}^{-1}\Bigl[ \frac{\Gamma\bigl( \sigma+\frac{j-s}{s+1} \bigr)}
                          {\Gamma\bigl( \sigma+\frac{2+j-s}{s} \bigr)}
  ;\; x\Bigr] = 
  \frac{x^{j-s} (1-x)^{\frac{1+s(1+j-s)}{s+1}}}
         {\Gamma\bigl(\frac{2(s+1)+js-s^2}{s+1}\bigr)} \; ,
\label{melb2}
\end{equation}
which for all $j=0,1,\dots, (s-1)$, is a positive function for $x\in (0,1)$.
Then Eq. (\ref{FStri}) can be viewed as a $(s-1)$--fold convolution of
positive functions, which by (\ref{mellconv}) is itself positive
for $x\in (0,K_s)$. The upper edge $K_s$ of the support 
can be read off from the prefactor in Eq.(\ref{FStri}).

In a similar way one can prove positivity of Raney distribution $W_{p,r}(x)$
for natural values of the parameters, provided $r \le p$.
Here is the streamlined version of the proof.
We use the aforementioned formula 8.4.2.3 of 
\cite{PBM98} for $b>a$, 
\begin{equation}
 {\cal M}^{-1} \Bigl[ \frac{\Gamma(\sigma+a)}{\Gamma(\sigma+b)} ; x\Bigr] =
\bigl[\Gamma(b-a)\bigr]^{-1} x^a (1-x)^{b-a-1},
\label{mellininv3}
\end{equation}
which describes a positive function for $0< x< 1$. We quote now the
full version of the analog of Eq.(\ref{FStri}) for $R_{p,r}(\sigma)$,
with $\sigma=n+1$ and $p=2,3,\dots \;$ ,
\begin{eqnarray}
R_{p,r}(\sigma) &= &  \frac{r}{\sqrt{2\pi}}  
\frac{p^{r-p-1/2}}{(p-1)^{r-p+3/2}}
 \Bigl[\frac{p^p}{(p-1)^{p-1}}\Bigr]^{\sigma} \times 
 \nonumber \\ 
 & \times & \; 
\frac{\Gamma\bigl(\sigma+\frac{r-p}{p}\bigr)}{\Gamma(\sigma)} 
 \; 
 \prod_{j=1}^{p-1} \Bigl[
\frac{\Gamma\bigl(\sigma+ \frac{r-p+j}{p}\bigr)}
     {\Gamma\bigl(\sigma+ \frac{r-p+j+1}{p-1}\bigr)}
 \Bigr] .
\label{Rantri}
\end{eqnarray}
Consider first the case $1\le r < p$. The weight function $W_{p,r}(x)$ is the 
inverse Mellin transform, $W_{p,r}(x)={\cal M}^{-1}\bigl[ R_{p,r}(\sigma); x \bigr]$,
and, from (\ref{Rantri}) via Eq.(\ref{mellconv}), it is the $p$--fold 
Mellin convolution of $\Gamma[\sigma+(r-p)/p]/\Gamma(\sigma)$
and of $p-1$ factors in the product in Eq.(\ref{Rantri}). 
For each of these individual ratios of gamma function  relation
(\ref{mellininv3}) holds. Therefore, with Eq.(\ref{Rantri}), the positivity
of the distribution $W_{p,r}(x)$ follows. 

In the second case, $r=p$, only $p-1$ factors in Eq.(\ref{Rantri}) intervene
in the convolution. Hence $W_{p,p}(x)$ is also positive.
If $r > p$ the first ratio in Eq.(\ref{Rantri})
destroys the positivity, so no further considerations are needed
to show that in this case the function $W_{p,r}(x)$ 
is not a probability distribution.

\section{Relations between Raney distributions $W_{p,r}(x)$}
\label{append3}

The Raney distribution $W_{p,r}(x)$ may be implicitly written 
by its $S$-transform, which allowed M{\l}otkowski \cite{Ml10} 
to establish a relation between distributions
with various values of their parameters in terms of the free 
multiplicative convolution denoted by ${\boxtimes}$
(compare Eq. 4.10 in \cite{Ml10}),  
\begin{equation}
W_{1+p,1} \; {\boxtimes} \; W_{1+q,1} = W_{1+p+q,1} . 
\label{frcon1}
\end{equation}
This result, valid for $p,q >0$,
can be generalized for an arbitrary positive $r$, 
 \begin{equation}
W_{p,r} \; {\boxtimes} \; W_{1+q,1} = W_{p+rq,r} . 
\label{frcon2}
\end{equation}
 Moreover, there exists yet another relation,
 \begin{equation}
[W_{1+p,1}]^{\boxtimes s} = W_{1+sp,1}
\label{frcon3}
\end{equation}
 which holds for $s>0$ and 
is equivalent to the multiplicative convolution property
for the Fuss--Catalan distribution.

\end{document}